\documentstyle[editedbook,epsfig,psfig,epsf]{mq}
\def\eg{{\it e.g.} }
\def\etal{{\em et al.} }
\def\ie{{\em i.e.} }

\def\msun{$M_{\odot}$ }

\def\cm2{cm$^2$ }
\def\se1{s$^{-1}$ }

\def\degree{$^{\circ}$ } 

\begin{opening}
\title{On the correlation between radio and X-ray flux in Low/Hard state Black
Holes} 
\author{E. Gallo$^1$, R. P. Fender$^{1}$, G. G. Pooley$^2$}
\institute{$^1$ Astronomical Institute ``Anton Pannekoek'' and Center for High
Energy Astrophysics, Kruislaan 403, 1098 SJ Amsterdam, the Netherlands.\\
$^2$ Mullard Radio Astronomy Observatory, Cavendish Laboratory, Madingley
Road, Cambridge CB3 0HE, UK.\\}
\end{opening}

\runningtitle{On the correlation between radio and X-ray flux in Low/Hard
state Black Holes}
\runningauthor{Gallo, Fender, Pooley}

\begin{document}
\vspace{-0.5cm}
\begin{abstract}
{\small Radio emission from X-ray binary systems (XRBs) has
developed in 
recent years from being peculiar phenomenon
to being recognised as an ubiquitous property of several classes of
XRBs. In this scenario the synchrotron emission is interpreted as the
radiative signature of jet-like 
outflows, some or all of which may possess relativistic bulk motion.
We have analysed a collection of quasi-simultaneous radio/X-ray observations
of  
Black Holes in the Low/Hard X-ray state, finding evidence
of a clear correlation between their fluxes over many orders of magnitude in
luminosity.  
Given that the correlation extends down to GX 339-4 and V404 Cyg  
in quiescence, we can confidently assert that even at accretion rates
as  
low as $\sim 10^{-5}$ $\dot{m}_{Edd}$ a powerful jet is being formed. 
The
normalisation of the correlation is very similar
across a sample of nine sources, implying that it is nearly independent of 
jet 
inclination angle. Remarkably, V 404 Cyg is the second source (after GX
339-4) to  
show the correlation 
$S_{radio}\propto S_{X}^{+0.7}$ from quiescent
level up to close to the High/Soft state transition. Moreover, 
assuming the same physics and 
accretion:outflow coupling for all of these systems, the simplest
interpretation for the 
observed scenario is that
outflows in 
Low/Hard state do not have large bulk Lorentz factors.} 
\end{abstract}
%**************************************************
\section{Introduction: radio emission from Low/Hard state Black Holes}
Radio emission is often observed from X-ray binaries, particularly transient
systems and especially Black Hole (BH) candidates (for
detailed reviews see \eg \cite{hh}, \cite{mr},
\cite{rob1}, \cite{rob2}). 
The Low/Hard state is one of the five `canonical' X-ray states observed from
BH X-ray binaries in our Galaxy. It is charaterized by a hard (spectral index
$\sim$ 1.5) X-ray spectrum, associated with a steady, self-absorbed jet which
emits  
synchrotron radiation; the Off state may simply be the Low state `turned down'
to lower accretion rates. X-ray spectra of High/Soft state
BHs     
are dominated by thermal radiation, while the radio emission drops below
detectable levels, probably corresponding to the physical disappearance of the
jet. For the so called Intermediate and Very High states the connection
between X-ray and  
radio properties is not yet completely established.
Transitions between 
states are often associated with multiple ejections of synchrotron emitting
material, possibly with high Lorentz factors.\\
There is an extremely strong correlation between radio and X-ray emission: the
former has been directly observed to arise in outflows and to produce
synchtrotron emission from a population of high energy electrons. The latter
has been inferred to arise via Comptonisation by a thermal electron
distribution. 
All the evidence points to the corona in these systems being physically
related to the presence of a jet: by far the simplest interpretation therefore
is that the Comptonising region is just the base of the relativistic outflow.

\section{Jet/corona coupling: $S_{radio}$ versus $S_{X}$ }

\begin{figure}[vtb]
\centering
\psfig{file=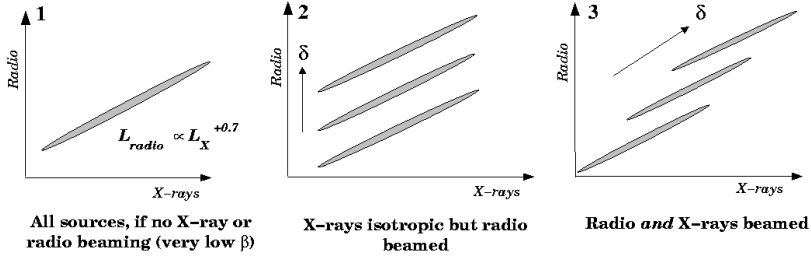,width=4cm,angle=-90}
\caption{Different placings are predicted for Low/Hard state BHs in the
$S_{radio}$ vs. $S_{X}$ plot in case of: no or very low beaming (left);
isotropic X-rays and beamed radio emission (centre); beamed radio and X-rays
(right).}  
\label{fig:scenario}
\end{figure}

By means of simultaneous radio/X-ray obsevations performed on the Galactic BH
GX 339-4, Corbel \etal (in 
preparation) have found an interesting correlation between fluxes, namely the 
radio 
density flux at 8.6 GHz scales as the 3-9 keV X-ray flux raised to 0.7.\\
Assuming the same physics and jet/corona coupling for all sources,
based on simple arguments one would predict the following
scenario in the $S_{radio}$ vs. $S_{X}$ parameter space (see sketch in fig. 
\ref{fig:scenario} for clarity):

\begin{figure}[vtb]
\centering
\psfig{file=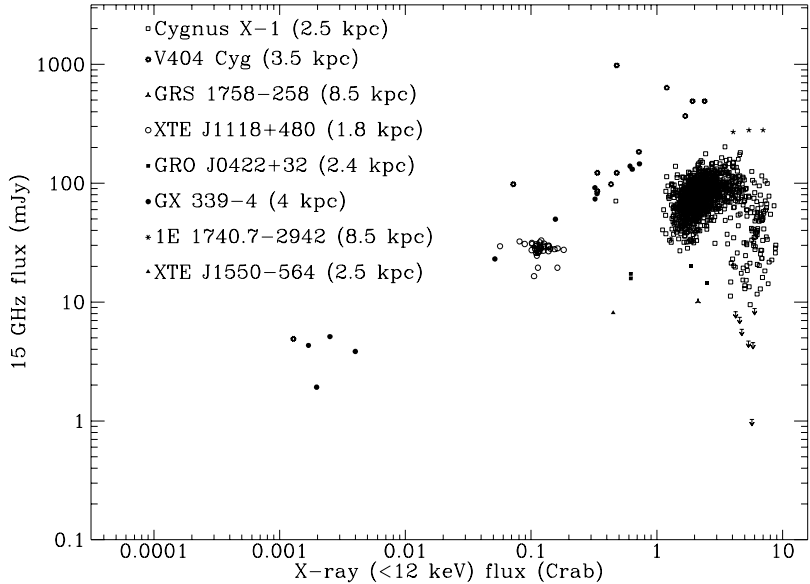,width=8.5cm,angle=-90}
\caption{Radio density flux (mJy) at 15 GHz is plotted against X-ray flux
(Crab)
below 12 keV  
for a sample 
of nine Low/Hard state BHs, scaled to a distance of 1 kpc. An evident
correlation 
between these two bands appears, and holds over more than three orders of
magnitude 
in accretion rate.}
\label{fig:correlation}
\end{figure}

\begin{enumerate}
\item All sources lying on a line with the same slope if neither X-ray or
radio emission were strongly 
beamed (\ie low $\beta$).
\item Different sources lying on lines with the same slope but different
normalisations 
if X-rays were isotropic but radio was beamed, with radio-brighter
sources corresponding to higher Doppler factors.
\item As point 2. but with higher Doppler factors sources being 
brighter in both radio and X-ray if either were beamed.
\end{enumerate} 

With the purpose of verifying if and possibly which kind of 
relation could be shared 
among a wide sample of sources in the same spectral state, we collected a  
variety of quasi-simultaneous radio/X-ray observations of Low
state BHs and plotted the radio vs. the X-ray
fluxes (below 12 keV) scaled to a distance of 1 kpc (units
are mJy 
vs. Crab). As shown in \ref{fig:scenario}, there is actually clear evidence for a strong
correlation, which - again - extends 
over more than three orders of magnitude in accretion rate, from around
$10^{-5}\dot m_{Edd}$ (calculated supposing $M_{BH}=$10 \msun)
up to close to the High/Soft state transition. 
It is important to stress that, in principle, the range of normalisation
values can be extremely large. Assuming again the same physics
for all 
sources, and that X-rays are isotropic while radio is beamed, the ratio
$S_{radio}/S_{X}$ for two sources would equal the ratio between their Doppler
factors (calculated as sum of approaching and receding jet) to the second
power. For instance, with $\beta$=0.998 (i.e. $\Gamma$=16) and angles ranging
between 1\degree - 90\degree, that ratio covers
12 orders of magnitude!  

%*************************************************************************

\section{Special cases: V 404 Cyg and GX 339-4}

\begin{figure}
\centering
\psfig{file=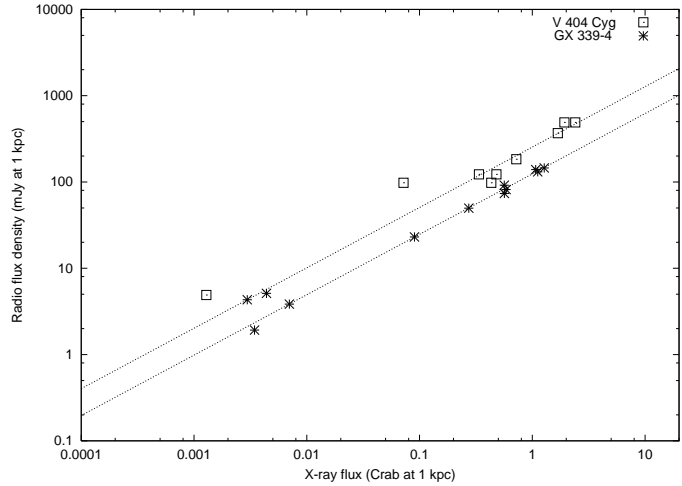,width=6.5cm,angle=-90}
\caption{For both V 404 Cyg and GX 339-4 (Corbel \etal, in preparation), from
quiescence up to 
High state transition, we found $S_{radio}= k \times S_{X}^{+0.7}$, with
$k_{V404}=254 \pm 18 $ and $k_{GX339}=124 \pm 2 $.}
\label{fig:fit}
\end{figure}

Remarkably, we found that the data of V 404 Cyg and GX 339-4, 
 \ie the sources for which we have the 
largest range in luminosity at our disposal,
are well fitted by a power law with the same slope, as shown in
fig. \ref{fig:fit}: we obtained $S_{radio}=k \times 
S_{X}^{+0.7\pm 0.1} $, with 
$k_{V404}=254 \pm 18 $ and $k_{GX339}=124 \pm 2 $. For the same simple
arguments we mentioned in the previous section, 
assuming isotropic X-rays and beamed radio emission, the ratio
$k_{V404}/k_{GX33}$ should be equal to
$(\delta_{V404}/\delta_{GX339})^2$.  
GX 339-4 seems to be at quite low inclination (
15\degree- 30\degree according with \cite{wu01} , \cite{soria}) while 
$i_{V404}$=56\degree (\cite{shab}, \cite{wag}); adopting these values  the
ratio 
$(\delta_{V404}/\delta_{GX339})^2 < 1$ for \emph{any} value of $\beta$=v/c.
Thus, V 404 being radio-brighter than
GX 339-4 despite the higher inclination excludes both the second and
the third scenario described in the previous section. The simplest possible
explanation requires very low beaming for both sources, attributing the
little 
scatter in normalisation to intrinsic errors.
Obviously a great uncertainty in this argument is introduced by distance
determinations, 
which strongly affect all the calculations. In addition we can not rule out
the possibility of a much more complex physical model behind this relation.

%***********************************************************************

\section{Summary and conclusions}

The results of our analysis can be summarized as follows:
\begin{itemize}
\item In Low/Hard state BHs the observed radio and X-ray fluxes are correlated
over more than three orders of magnitude in accretion rate, implying a strong
jet/corona coupling; no lower limit to the relation has been found.
\item We can confidently assert that, even at accretion rates as low as $\sim$
$10^{-5} \dot m_{Edd}$ a powerful jet is being formed.
\item V 404 Cyg is the second source to display $S_{radio} \propto
S_{X}^{+0.7}$, 
from quiescence up to Soft state transition. A physical explanation for this
relation is proposed by Markoff \etal (\cite{sera}).
\item Above $\sim 10^{-2} \dot m_{Edd}$ the jet disappears within a factor of a
few, probably in all sources (observed in three sources and no exceptions)
\item Comparison of different sources may indicate that jets in Low/Hard have
a low 
velocity compared to those in transient outbursts. 
Of course better distance and inclination determinations are strongly required
to probe this conjecture.\\
Finally, we would like to mention that the possible detection of the Soft
X-ray Transient A0620-00 at the predicted -- extrapolating the relation we
found -- radio level would demonstrate that jet production is an ubiquitous
mechanism between $10^{-7}$ and $10^{-2} \dot m_{Edd}$.
\end{itemize}


\begin{thebibliography}{}
\bibitem{hh}
Hjellming R. M., Han X. H., 1995, 'Radio properties of X-ray binaries', in
Lewin W. H. G., van Paradijs J., van den Heuvel E. P. J., eds, \emph{X-ray
binaries}, CUP, 308.

\bibitem{mr}
Mirabel, I. F., Rodriguez, L. F.,1999, ARA\&A, {\bf 37}, 443.

\bibitem{rob1}
Fender, R., 2001, Astrophysics and Space Science, {\bf 276} (suppl), 69.
 
\bibitem{rob2}
Fender, R., 2001, MNRAS, {\bf 322}, 31.

\bibitem{wu01}
Wu, K., Soria R., Hunstead R. W., Johnston H. M., 2001, MNRAS, {\bf 320},
177. \
 
\bibitem{soria}
Soria R., Wu K., Johnston H. M., 1999, MNRAS , {\bf 310}, 71.


\bibitem{shab}
Shahbaz, T., Ringwald, F. A., Bunn, J. C., Naylor, T., Charles, P. A.,
Casares, J., 1994,  MNRAS, {\bf 271}, L10. 

\bibitem{wag}
Wagner, R. M., Kreidl, T. J., Howell, S. B., Starrfield, S. G., 1992, 
ApJ, {\bf 401}, L97. 

\bibitem{sera}
Markoff, S., Nowak, M., Corbel, S., Fender, R., Falcke, H., 2002, A\&A
\emph{submitted}. 


\end{thebibliography}
\end{document}